\documentclass[pdftex,twocolumn,epjc3]{svjour3}          
\RequirePackage[T1]{fontenc}

\smartqed  

\RequirePackage{graphicx}
\RequirePackage{mathptmx}      
\RequirePackage{flushend}
\RequirePackage[numbers,sort&compress]{natbib}
\RequirePackage[colorlinks,citecolor=blue,urlcolor=blue,linkcolor=blue]{hyperref}
\usepackage{amsmath, amsfonts}
\usepackage{amssymb, color}

\journalname{Eur. Phys. J. B}

\newcommand{\beqa}{\begin{eqnarray}}
\newcommand{\eeqa}{\end{eqnarray}}

\newcommand{\rarr}{\rightarrow}

\begin{document}

\title{Simple analytical model of a thermal diode}


\author{Saurabh Kaushik\thanksref{addr1, addr2}
        \and
        Sachin Kaushik\thanksref{addr1, addr3}
        \and
        Rahul Marathe\thanksref{e1, addr1}
}

\thankstext{e1}{e-mail: maratherahul@physics.iitd.ac.in}

\institute{Department of Physics, Indian Institute of Technology, Delhi, Hauz Khas 110016, 
New Delhi, India. \label{addr1}
\and
Soft Condensed Matter Group, Raman Research Institute, C. V. Raman Avenue, Sadashivanagar, 
Bangalore 560080, India. \label{addr2}
\and
Theoretical Sciences Unit, Jawaharlal Nehru Centre for Advanced Scientific Research,
Jakkur, Bangalore 560064, India. \label{addr3}
}

\date{Received: date / Accepted: date}

\maketitle

\begin{abstract}
Recently there is a lot of attention given to manipulation of heat by constructing thermal devices 
such as thermal diodes, transistors and logic gates. Many of the models proposed have an asymmetry 
which leads to the desired effect. Presence of non-linear interactions among the particles is also 
essential. But, such models lack analytical understanding. Here we propose a simple, analytically 
solvable model of a thermal diode. Our model consists of classical spins in contact with multiple 
heat baths and constant external magnetic fields. Interestingly the magnetic field is the only 
parameter required to get the effect of heat rectification.

\end{abstract}

\section{Introduction} Recently heat transport in microscopic systems has attracted a lot of attention. These studies mainly involve studying heat transport in terms of verification of Fourier's law at molecular scale using detailed microscopic models \cite{A. Dhar,Lepri et.al,Saito,B. Hu et.al,Chang2008 et.al,Yang2010 et.al}. Looking at the working of small scale devices, such as engines, thermal/electronic pumps or analogous thermal rectification devices like diode, transistors, logic gates etc. is also of interest \cite{Lo et.al,Saira et.al,Baowin2006a,BaowinRMP2012,Baowin2004,Liang et.al,Li et.al,Wang2007 & Li, Marathe07}. These effects have been studied in classical as well as quantum mechanical systems, and have also been realized experimentally \cite{Chang2006 et.al,Kim et.al,Wu2007 & Li,Wu2008 & Li,Yang2008 et.al,Yang2009 et.al,Hu2009 et.al,Gonzalez et.al,Jiang et.al, ChenNatComm17}. Thermal diodes, transistors and logic gates have been modeled using classical harmonic oscillator chains with non-linear interactions and acted upon by different on-site potentials \cite{Baowin2006a,BaowinRMP2012, Wang2007 & Li, Baowin2004}. The mismatch of the on-site potentials and phonon modes as well as the non-linear interactions play crucial role in such models to get the desired effect of heat rectification. These thermal devices were also modeled using classical spins \cite{BLeePRE11, Bagchi13, Bagchi15}. In these models anisotropic parameter and mismatch of the spin flipping rate at the interface of two spin networks in contact with different heat baths are the underlying factors for the rectification of heat current. 

Due to non-linear interactions among the constituents analytical understanding becomes difficult. However, many interesting analytical attempts have been made to
understand basic principles behind heat rectification, for example in harmonic chains with quartic potential \cite{PereiraPRE10} as well as studying sufficiency conditions for thermal rectification  \cite{PereiraPRE11}. In mass-graded harmonic chains with temperature dependent effective potential \cite{PereiraPRE17} and also in quantum harmonic chains 
\cite{PereiraPLet10}. Here we propose a simple model of a thermal diode which is in principle analytically tractable. Our model consists of coupled classical spins, with nearest neighbor interactions. One segment of the spins is in contact with a thermal reservoir at temperature $T_L$ and the other with a thermal reservoir at temperature $T_R$. Spins are also acted upon by external constant magnetic fields $h_L$ and $h_R$ respectively. Interestingly the externally applied magnetic field is the only control parameter for our thermal device. 

The paper is organized as follows. We first describe the model of the thermal diode then we provide 
the results for this model. We conclude with a discussion of our results. 

\section{Model}
\begin{figure}[ht]
\centering
\includegraphics[width=1.0\columnwidth]{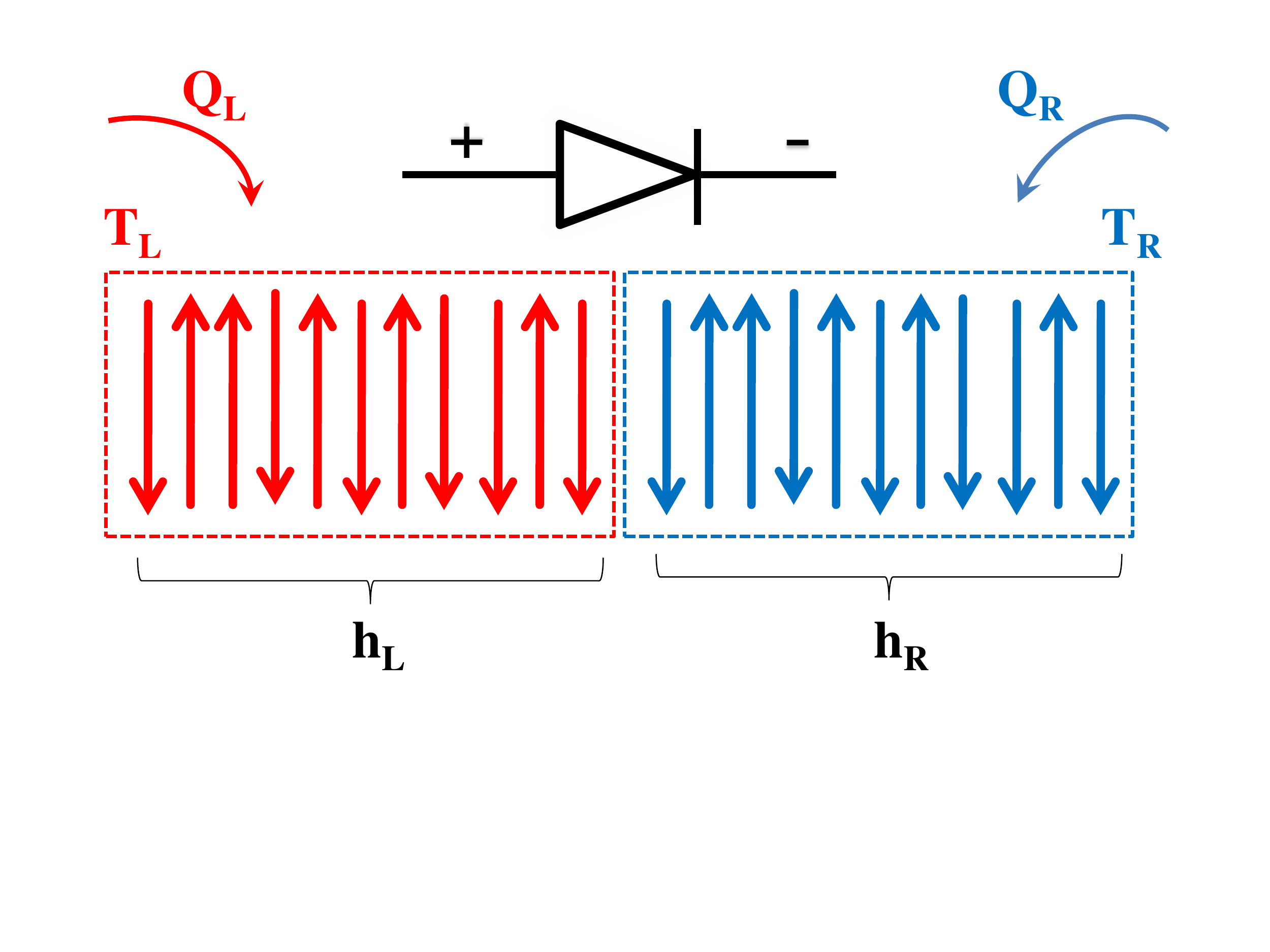}
\vspace{-2.5cm}
\caption{Schematic diagram of thermal diode with classical Ising spins. The red (left) segment consisting $N_L$ spins subjected to an external magnetic field $h_L$ and in contact with heat bath at temperature $T_L$. The $N_R$ spins in blue (right) segment are in contact with heat bath at temperature $T_R$ acted upon by a constant external magnetic field $h_R$.} 
\label{fig1}
\end{figure}

The system we consider is divided into left and right segments.
The left (suffix $(L)$) segment has $N_L$ number of Ising spins and right (suffix $R$) 
segment has $N_R$ number of Ising spins as shown in Fig. \ref{fig1}. The Hamiltonian of the system 
is given by:
\beqa
\mathcal{H} = -J\sum\limits_{\langle i,j \rangle} \sigma_i ~\sigma_j -\mu h_L\sum\limits_{i=1}^{N_L} \sigma_i -\mu h_R\sum\limits_{i=1}^{N_R} \sigma_i ,
\label{Hamilt}
\eeqa

where, $J$ is the interaction energy between the spins, $\mu$ is the magnetic moment of spins, $h_{L}$ and $h_{R}$ are the external magnetic fields. Each spin can take two values $\sigma_i$=$\pm1$
and $\langle i,j\rangle$ represents nearest neighbor interaction. Since the spins are in thermal contact with the temperature reservoirs, the flipping of the spins is a stochastic process and the master equation governing this stochastic evolution of probabilities is:
\beqa
\frac{\partial \hat{P}(\left\{ \sigma \right\},t )}{\partial t}= \hat{\mathcal{T}}
\hat{P}(\left\{ \sigma \right\},t),
\label{Mastereq}
\eeqa
where $\hat{\mathcal{T}}$ is the Transition Matrix, $\hat{P}(\left\{ \sigma \right\},t)$ is the spin distribution function, where $\left\{ \sigma \right\}$ represents spin 
configuration at time $t$, such that $\hat{P}(\{\sigma\}, t)=\hat{P}(\sigma_1, \sigma_2,...,t)$. 

The dynamics of the spins with respective heat baths is modeled by a Metropolis algorithm generalized 
to accommodate multiple heat reservoirs. The choice of algorithm is generic and our results 
qualitatively do not depend on particular form of the flipping probabilities, for example
Glauber dynamics. The elements in the transition matrix $\hat{\mathcal{T}}$ 
give the rates of transition from one configuration to other. If a spin flips then the rate of 
this flip is given by:
\beqa
r^{L,R}_{\sigma_i}=~min(1,~e^{-\beta_{L,R} \Delta{E_{L,R}}}),
\label{fliprt}
\eeqa
where, $\beta_{L,R}=1/k_BT_{L,R}$ is chosen depending on which segment the spin belongs to and 
$\Delta{E_{L,R}}=2J(\sigma_i \sigma_{i-1} + \sigma_i \sigma_{i+1}) + 2\mu h_{L,R} \sigma_i$ 
is the difference in the energy after and before the flip. In our analysis we have chosen 
the Boltzmann constant $k_B$ and $\mu$ to be unity. For larger system sizes one may also resort to 
Monte Carlo simulations where a spin is chosen at random and is flipped with probabilities given 
above. Here the flipping probability is nothing but the rate $r^{L,R}_{\sigma_i}$ multiplied by the 
the time step $dt$. In each $dt$ time step only a single spin flip is allowed.

This modified Metropolis algorithm for multiple heat baths allow us to write general expression for 
the heat currents in the system. $\dot{Q}_{L,R}$ are the heats coming from the left and the right baths
respectively. In our model, heat coming into the system from the reservoirs is taken to be positive. 
The expression for the heat currents are:
\beqa
\dot{Q}_{L,R}=\sum_{\left\{\sigma\right\}} r^{L,R}_{\sigma_i}  \Delta{E_{L,R}}  
\hat{P}(\left\{ \sigma \right\}).
\label{heatcurr}
\eeqa
In principle for any system size, given the transition matrix $\hat{\mathcal{T}}$,  
Eq. (\ref{Mastereq})  can be solved analytically in the steady state. Since in the steady state 
$\frac{\partial \hat{P}}{\partial t}=0$, thus $\hat{P}(\left\{ \sigma \right\})$ is nothing but the 
eigenvector of the matrix $\hat{\mathcal{T}}$ corresponding to the eigenvalue zero \cite{vankampen}.
Once the steady state probabilities $\hat{P}(\left\{ \sigma \right\})$ are obtained, both heat 
currents can also be evaluated. In the next section we describe the working of the thermal diode
using definitions above. 

\section{Thermal Diode} We now describe working of our model as a thermal diode. Here we give a particular example for a small system consisting of just two Ising spins, first one in contact with a bath at temperature $T_L$ and second with $T_R$. We fix $h_R=0$ without loss of generality. 
With these parameters and Eq. (\ref{fliprt}) we can determine the matrix elements of 
$\hat{\mathcal{T}}$. Using steady state probabilities P($\sigma_1,\sigma_2$) along with the 
normalization condition $\sum\limits_{\sigma_1, \sigma_2}P(\sigma_1,\sigma_2)=1$, heat currents 
defined in Eq. (\ref{heatcurr}) turn out to be:

\beqa
&\dot{Q}_L& =~2(J+\mu h_L)~\left[e^{-2\beta_L(J+\mu h_L)}P(1,1)-P(-1,1)\right]  \nonumber \\
&&+2(\mu h_L-J) \left[e^{-2\beta_L(\mu h_L-J)}P(1,-1)-P(-1,-1)\right], \nonumber \\ 
&\dot{Q}_R&=~2J~\left[e^{-2\beta_RJ}P(1,1)-P(1,-1) -P(-1,1)+e^{-2\beta_RJ}P(-1,-1)\right]. \nonumber \\ 
\label{qlqr}
\eeqa
where,
\beqa
&P(1,1)&=~e^{2 \beta_L \mu h_L+\beta_R J}~ [e^{\beta_L (J-\mu h_L)} \cosh(\beta_L (\mu h_L-J)+\beta_R J) \nonumber \\
&&+\cosh (\beta_R J)]/\mathcal{D},\nonumber \\
&P(1,-1)&=~e^{2 \beta_L \mu h_L-\beta_R J}~ [e^{2 \beta_R J-\beta_L (\mu h_L+J)} 
\cosh (\beta_L (\mu h_L+J)-\beta_R J)  \nonumber \\ 
&&+\cosh (\beta_R J)]/\mathcal{D}, \nonumber \\
&P(-1,1)&=~[e^{-\beta_L (J+\mu h_L)} \cosh (\beta_L (\mu h_L-J))   \nonumber \\
&&+~\cosh (2 (\beta_L-\beta_R) J)]/\mathcal{D}, \nonumber \\
&P(-1,-1)&=~e^{\beta_R J} ~[e^{\beta_L (J-\mu h_L)} \cosh (\beta_L(\mu h_L+J)-\beta_R J) \nonumber \\
&&+~\cosh ((2 \beta_L-\beta_R)J)]/\mathcal{D},\nonumber \\
&\mathcal{D}&=~2\cosh (\beta_R J) [e^{2 \beta_L \mu h_L} \cosh (\beta_R J)+
e^{\beta_R J} \cosh (2\beta_L \mu h_L)] \nonumber \\
&&+~2 \left[e^{\beta_R J}+\cosh (\beta_R J)\right] \cosh ((2 \beta_L-\beta_R) J), \nonumber
\eeqa

Eq. (\ref{qlqr}) is the steady state expression for $\dot{Q}_L $ and $\dot{Q}_R$ in terms of 
$P(\sigma_1,\sigma_2)$. After some simplification we get:
\beqa
&\dot{Q}_L&=~4~J \sinh (2 (\beta_R-\beta_L) J)/\mathcal{D},\nonumber\\
&\dot{Q}_R&=~4~J \sinh (2 (\beta_L-\beta_R) J)/\mathcal{D}.
\label{qlqr_exp}
\eeqa

\begin{figure*}[!tbhp]
\centering
\includegraphics[width=1.6\columnwidth]{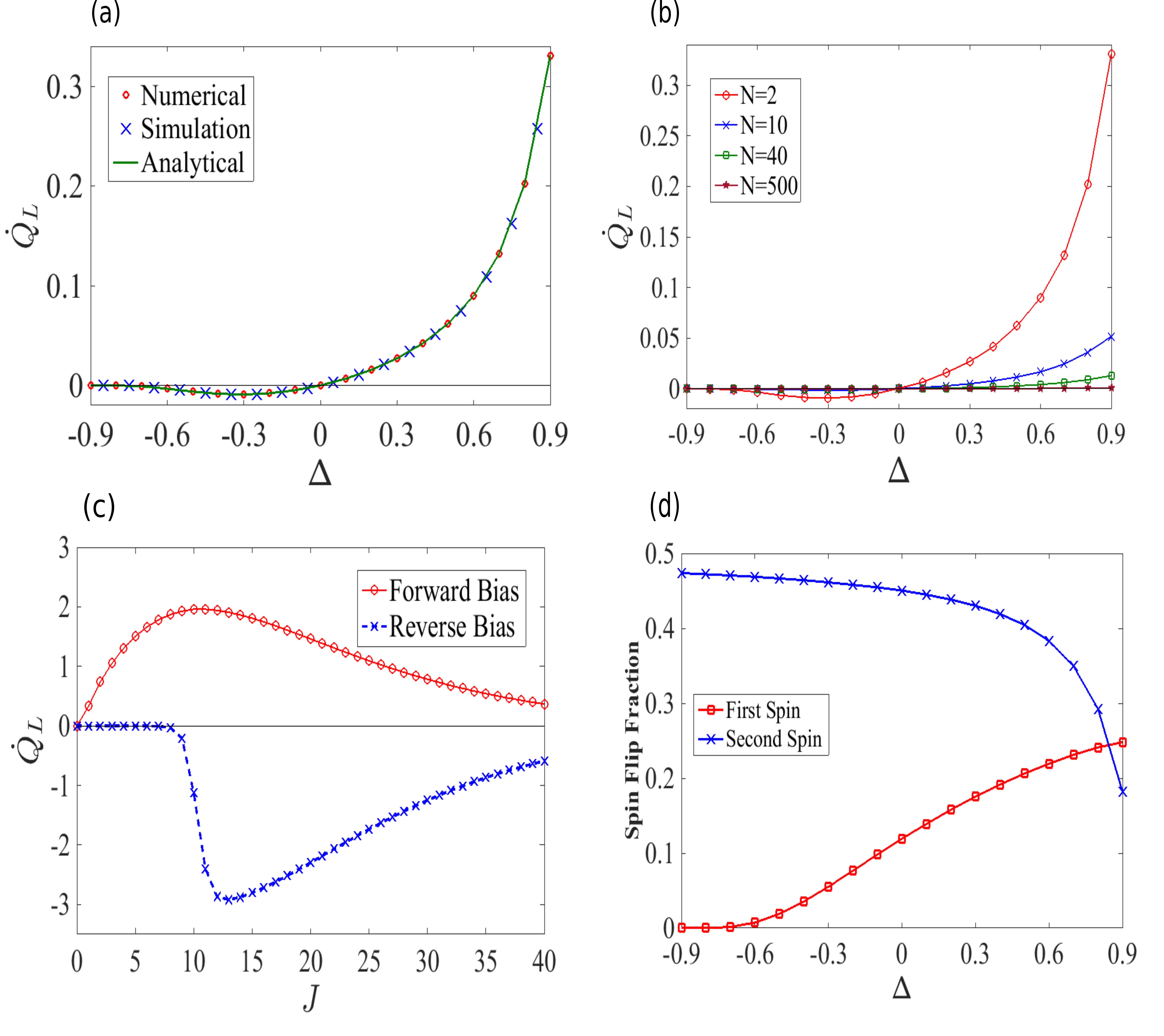}
\caption{The parameters used are $J=1$, $h_L=10$, $T_0=10$, and $h_R=0$. $(a)$ Heat current versus the temperature bias for the system with two Ising spins. The plot consists the results from numerical techniques, simulations and analytical solution of the Master Equation. $(b)$ Heat current versus the temperature bias for different system sizes. $(c)$ Heat current versus the interaction term $J$. Here $h_L=10$ and $h_R=0$. For forward biased case $T_L=19$ and $T_R=1$, and for the reverse biased case $T_L=1$ and $T_R=19$. $(d)$ Spin flip fraction versus the temperature bias for the system with two Ising spins. The plot shows the fraction of first spin and the second spin flips in a simulation. Zero line is 
just a guide for the eyes.} 
\label{fig2}
\end{figure*}

\begin{figure}[ht]
\hspace{-1cm}
\centering
\includegraphics[width=1.\columnwidth]{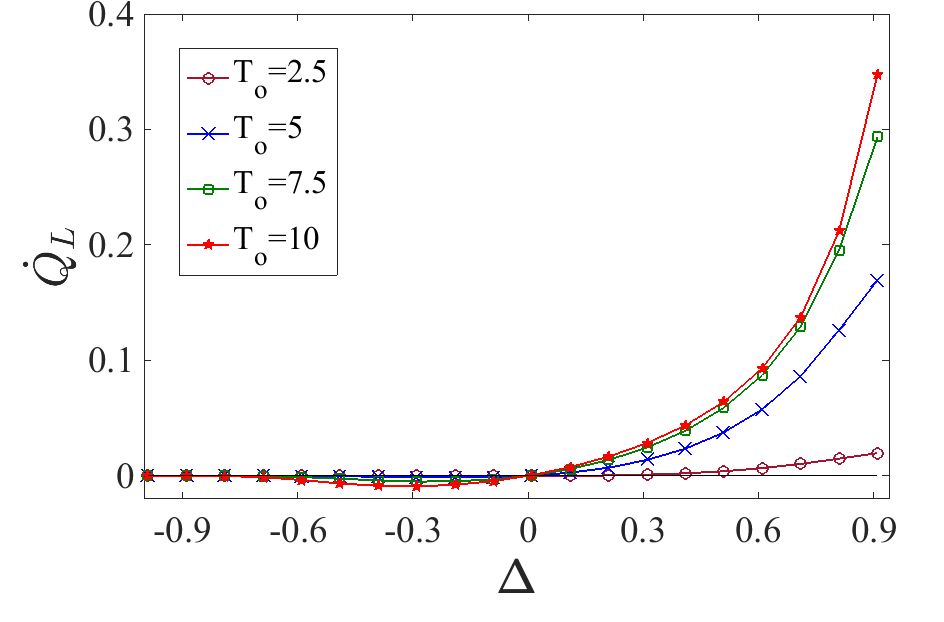}
\caption{Plot of  heat current versus $\Delta$ for different reference temperatures $T_0$ for a two spin system with $J=1$, 
$h_L=10$, $h_R=0$.}
\label{fig3}
\end{figure}
The heat currents $\dot{Q}_L$ and $\dot{Q}_R$ are related by just a sign change. 
Also the heat currents vanish when $T_L=T_R$, as expected. 

To study the characteristic of working of a diode, we define, $T_L=T_0(1+\Delta)$ and 
$T_R=T_0(1-\Delta)$ where $\Delta$ is relative temperature bias and $T_0$ is the reference 
temperature. Negative values of $\Delta$ correspond to reverse biased operation 
$(T_R >T_L)$ and positive to forward biased operation $(T_L>T_R)$. We plot the heat current 
$\dot{Q}_L$ as a function of $\Delta$. In Fig. \ref{fig2}$(a)$ the graph shows comparison 
of results obtained from direct simulations, numerical solution of the Master 
equation and exact solution Eq. (\ref{qlqr_exp}). Data from all the three calculations match 
perfectly. We can see that in the reverse biased case thermal current is extremely low as compared to 
the forward biased case. In Fig. \ref{fig2}$(b)$ we show the results for systems with different 
system sizes. The graphs again depict that our model works as a thermal diode even for larger systems. 
It is well known that the asymmetry in the system causes rectification. In the model considered 
here, the unequal magnetic fields on the two segments provide this asymmetry, making it work as 
a heat rectifying device. 

From Fig. \ref{fig2}$(c)$ it can be observed that the heat rectification occurs only when 
$J$ is less than $\mu h_L$. This happens because in the forward biased mode 
$T_L>T_R$ and though the spin experiences magnetic field $h_L$, thermal energy dominates and  
the left spin flips, resulting in large current from left to right bath. However 
in the reverse biased case magnetic field $h_L$ dominates over the interaction energy $J$ and 
since $T_L < T_R$, thermal energy is not enough to flip the spin. Hence, the current reduces 
drastically. To quantify this, in Fig. \ref{fig2}$(d)$ we plot the fraction of number of flips of 
the first and second spin as a function of $\Delta$. In the reverse biased mode number of spin 
flips of the first spin are negligibly small as compared to the second spin, resulting in small 
current. But as we go from reverse biased mode to forward biased mode they slowly become 
comparable to each other and a large current flows in the system. We can also obtain asymptotic 
values of $\dot{Q}_{L,R}$ when $\Delta \rarr \pm 1$. Notice that as $\Delta \rarr 1$, $T_L\rarr 2T_0$ 
and $T_R\rarr 0$ hence terms with $\beta_R$ dominate conversely when $\Delta \rarr -1$, $T_L\rarr 0$ 
and $T_R\rarr 2T_0$ and terms with $\beta_L$ dominate. From the Eqs. (\ref{qlqr}) and 
(\ref{qlqr_exp}) some simple algebra results into:
\beqa
\dot{Q}_L \rarr
\begin{cases}
4J/A & ,\Delta \rarr 1 \\ \\
{\large\frac{-2J \exp\left(-2\beta_R J\right)}{\left(B \exp\left(2\beta_L(\mu h_L-J) \right) 
~+~C\right)}\ }& ,\Delta \rarr -1, 
\end{cases}
\eeqa
where: 
\beqa
A&=& 2 \exp(2\beta_L(\mu h_L+J))+\exp(2\beta_L(J-\mu h_L))+3, \nonumber \\
B&=& \cosh(\beta_R J) \left(2 \cosh(\beta_R J) + \exp(\beta_R J)\right),\nonumber \\
C&=& 1 + \exp(-\beta_R J) \cosh(\beta_R J).\nonumber 
\eeqa 
It is straight forward to check that in the forward biased mode current 
approaches a finite positive value and in reverse biased mode, only when $\mu h_L > J$, current 
approaches zero from below as seen from Fig. \ref{fig2}$(a)$, $(c)$. For particular values of 
the parameters namely $\Delta=-0.9$ and $T_0=10$, we have $T_L=1.0$ and $T_R=19.0$, 
with $h_L=10.0$ and $h_R=0$, we get $\dot{Q}_L\sim 10^{-10}$ and for $\Delta=0.9$ it is $\sim0.35$. 
Fig.~\ref{fig3} shows the diode operation for different reference temperatures $T_0$. 

\begin{figure}[ht]
\centering
\includegraphics[width=1.0\columnwidth]{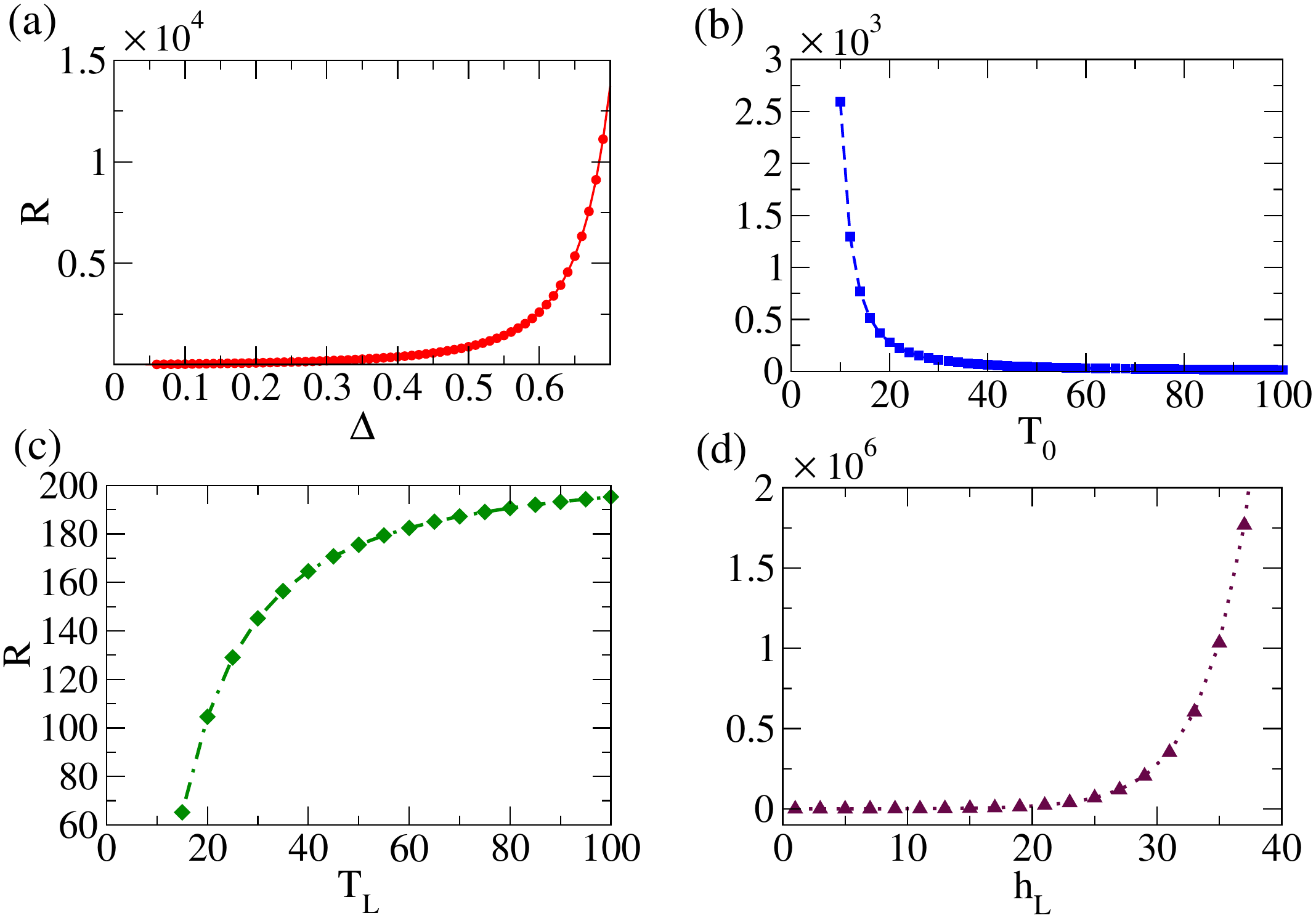}
\caption{Plot of rectification factor $R$ versus different model parameters. 
Parameters wherever applicable are $J=1$, $h_L=10$. $(a)$ $R$ versus 
$\Delta$ here $T_0=10$. $(b)$ $R$ versus $T_0$ with $\Delta=\pm 0.6$ in forward/reverse biased mode. 
Here $h_L$ may also be increased with $T_0$, so that $R$ does not decrease. $(c)$ $R$ versus $T_L$ 
where $T_R=10$ is fixed and $T_L$ varied. $(d)$ $R$ versus $h_L$ parameters used are 
$T_0=10$, $\Delta=\pm 0.5$ in forward/reverse biased mode.}
\label{fig4}
\end{figure}

Finally we study the rectification efficiency of our device. For this we define the rectification
factor $R$:
\beqa
R = \left(\frac{\dot{Q}_L-\dot{Q}_R}{\dot{Q}_R}\ \right)\times 100\% ,
\eeqa
where $\dot{Q}_L$ is measured when the device is in the forward biased mode and $\dot{Q}_R$ when it is 
in the reverse biased mode. Recent study has shown that this factor can be made independent of the 
system size with a ballistic spacer placed between two anharmonic chains \cite{CasatiArxv17}. 
We plot rectification factor for our model for different parameters in Fig. \ref{fig4}.
We can clearly see that for almost all parameters $R$ can be made very large. 

In heat conduction problems involving linear harmonic oscillators, heat current turns out to be 
proportional to the temperature difference $(T_L-T_R)$ \cite{A. Dhar, Marathe07}, for such a system 
rectification is not possible unless the interactions are non-linear \cite{BAgarwalla}.
In our model however the current does not depend linearly on the temperature difference as seen from 
Eq. (\ref{qlqr_exp}). Non-linearity is introduced through the flipping rates 
Eq. (\ref{fliprt}). Even then the analytical expressions for the heat currents are obtained, 
this distinguishes our model from the earlier studied models. 

\begin{figure}[ht]
\centering
\includegraphics[width=1.\columnwidth]{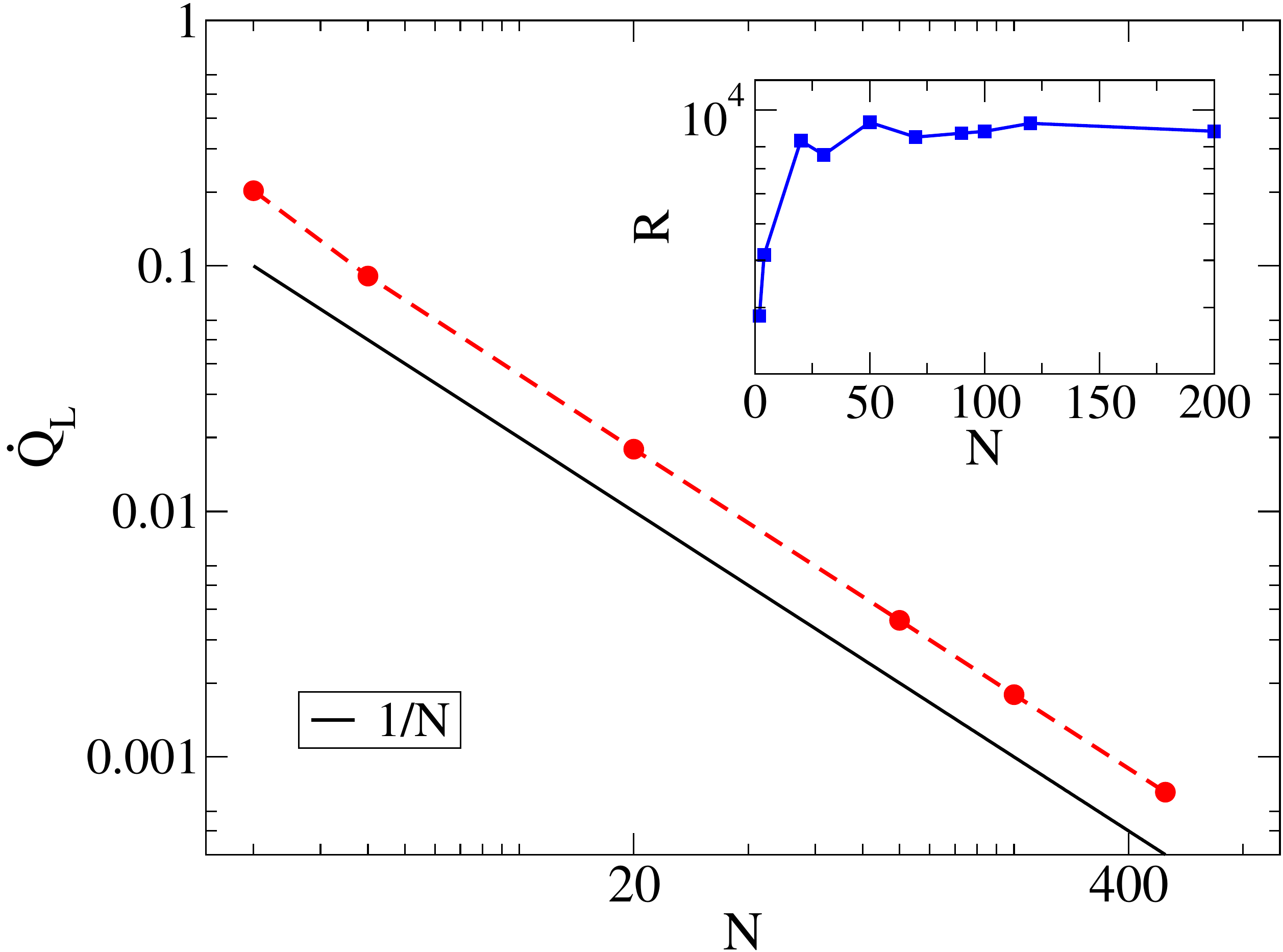}
\caption{Plot of heat current $\dot{Q}_L$ in the forward biased mode $(\Delta=0.8)$ as a function of system size $N$. One can clearly see the $1/N$ dependence
of heat current on system size. Inset rectification factor $R$ defined in the text as a function of system size for $|\Delta |=0.8$. Interestingly $R$ goes to 
a constant value as $N$ increases. Other parameters are $J=1$, $h_L=10$, $T_0=10$.}
\label{fig5}
\end{figure}

\section{Conclusion} 
To conclude, we have studied a simple model of a thermal diode composed of classical Ising spins 
connected to multiple heat reservoirs and are in the presence of constant external magnetic fields. 
Ising spins undergo stochastic dynamics governed by usual Metropolis algorithm, modified to take 
care of multiple heat baths. Interestingly external magnetic field is the only parameter which makes 
the models work as a thermal rectifier. Although our model has a non-linearity inbuilt in the 
transition rates, we are able to get analytical results for any system size. Here we note that, 
if say $N_L$ number of spins are in contact with reservoir at temperature $T_L$ and $N_R$ with 
temperature $T_R$, the current in large system size just scales as $\sim 1/N$ if $N_L=N_R=N$ $($Fig. \ref{fig5}$)$.
On the other hand rectification factor $R$ seems to become constant as system size increases $($inset of Fig. \ref{fig5}$)$,
which is an interesting effect. A similar two-level quantum mechanical model was studied in \cite{BLeePRB09} where the heat rectification was observed. 
Our study thus ascertains that even in classical discrete systems such a phenomenon is possible. 
Study on effect of long range interactions in our model will be interesting. 
Such interactions seem to increase the rectification efficiency \cite{AvilaPRE13}. 
However, analytical calculation of heat currents with long range interactions may not be possible. 
Extension of our model to other thermal devices namely transistors, logic gates is 
currently underway. We believe experimental realization of such a model is possible in nanoscale 
solid state devices. \\


\noindent {\bf Acknowledgments}\\
Authors thank the IIT Delhi HPC facility for computational resources. Authors also thank
Arnab Saha for careful reading of the manuscript. \\

\noindent{\bf Author contribution statement}\\
SauK and SacK performed simulations and analytical calculations. RM devised the study, performed simulations and analytical calculations.
SauK, SacK and RM wrote the paper.

\end{document}